\documentclass[11pt,draftclsnofoot,journal,onecolumn]{IEEEtran}
\usepackage{graphicx}
\usepackage{amsmath}
\usepackage{array}
\newcommand{\PreserveBackslash}[1]{\let\temp=\\#1\let\\=\temp}
\newcolumntype{C}[1]{>{\PreserveBackslash\centering}p{#1}}
\newcolumntype{R}[1]{>{\PreserveBackslash\raggedleft}p{#1}}
\newcolumntype{L}[1]{>{\PreserveBackslash\raggedright}p{#1}}
\usepackage[usenames]{color}
\usepackage{colortbl,booktabs}
\usepackage{tabularx}
\usepackage{multicol}
\usepackage{booktabs}
\usepackage[Symbol]{upgreek}
\usepackage{subfigure}
\usepackage{bm}
\usepackage{cite}
\usepackage{balance}
\usepackage[boxed]{algorithm2e}

\usepackage{threeparttable}
\usepackage{amsmath}
\usepackage{dcolumn}
\usepackage{multirow}
\usepackage{booktabs}
\usepackage{amsfonts}
\newcolumntype{d}[1]{D{.}{.}{#1}}

\begin{document}

\bibliographystyle{IEEEtran} 
\title{Turbo-Like Beamforming Based on Tabu Search Algorithm for  Millimeter-Wave Massive MIMO Systems}

\author{Xinyu Gao, \emph{Student Member}, \emph{IEEE}, Linglong Dai, \emph{Senior Member}, \emph{IEEE}, \\ Chau Yuen,  \emph{Senior Member}, \emph{IEEE}, and Zhaocheng Wang, \emph{Senior Member}, \emph{IEEE}

\vspace*{-3mm}
\thanks{X. Gao, L. Dai, and Z. Wang are with the Tsinghua National Laboratory
for Information Science and Technology (TNList), Department of Electronic Engineering, Beijing 100084, China (e-mail: gxy1231992@sina.com, \{daill, zcwang\}@tsinghua.edu.cn).}
\thanks{C. Yuen is with the SUTD-MIT International Design Center, Singapore University of Technology and Design, 20 Dover Drive, Singapore 138682, Singapore (e-mail: yuenchau@sutd.edu.sg).}
\thanks{This work was supported by National Key Basic Research Program of China (Grant No. 2013CB329203), National High Technology Research and Development Program of China (Grant No. 2014AA01A704), and National Nature Science Foundation of China (Grant Nos. 61271266 and 61201185).}}

\maketitle
\vspace*{-15mm}
\begin{abstract}
\vspace*{-3mm}
For millimeter-wave (mmWave) massive MIMO systems, the codebook-based analog beamforming (including transmit precoding and receive combining) is usually used to compensate the severe attenuation of mmWave signals. However, conventional beamforming schemes involve complicated search among pre-defined codebooks to find out the optimal pair of analog precoder and analog combiner. To solve this problem, by exploring the idea of turbo equalizer together with tabu search (TS) algorithm, we propose a Turbo-like beamforming scheme based on TS, which is called Turbo-TS beamforming in this paper, to achieve the near-optimal performance with low complexity. Specifically, the proposed Turbo-TS beamforming scheme is composed of the following two key components: 1) Based on the iterative information exchange between the base station and the user, we design a Turbo-like joint search scheme to find out the near-optimal pair of analog precoder and analog combiner; 2)  Inspired by the idea of TS algorithm developed in artificial intelligence, we propose a TS-based precoding/combining scheme to intelligently search the best precoder/combiner in each iteration of Turbo-like joint search with low complexity. Analysis shows that the proposed Turbo-TS beamforming can considerably reduce the searching complexity, and simulation results verify that it can achieve the near-optimal performance.
\end{abstract}

\vspace*{-6mm}

\section{Introduction}\label{S1}

\IEEEPARstart The integration of  millimeter-wave (mmWave) and massive multiple-input multiple-output (MIMO) is regarded as a promising technique for future 5G wireless communication systems~\cite{roh2014millimeter}, since it can provide orders of magnitude increase both in the available bandwidth and the spectral efficiency~\cite{marzetta10}. On one hand, the very short wavelength associated with mmWave enables a large antenna array to be easily installed in a small physical dimension~\cite{han2015large}. On the other hand, the large antenna array in massive MIMO can provide a sufficient antenna gain to compensate the severe attenuation of mmWave signals due to path loss, oxygen absorption, and rainfall effect~\cite{roh2014millimeter}, as the beamforming (including transmit precoding and receive combining) technique can concentrate the signal in a narrow beam.

MmWave massive MIMO systems usually perform beamforming in the analog domain, where the transmitted signals or received signals are only controlled by the analog phase shifter (PS) network with low hardware cost~\cite{roh2014millimeter}. Compared with traditional digital beamforming, analog beamforming can decrease the required number of expensive radio frequency (RF) chains at both the base station (BS) and users, which is crucial to reduce the energy consumption and hardware complexity of mmWave massive MIMO systems~\cite{bogale2014beamforming}.
Existing dominant analog beamforming schemes can be generally divided into two categories, i.e., the non-codebook beamforming and the codebook-based beamforming. For the non-codebook beamforming, there are already some excellent schemes. In~\cite{zhang2005variable,zhang2014achieving,bogale2014hybrid}, a low-complexity analog beamforming, where two PSs are employed for each entry of the beamforming matrix, is proposed to achieve the optimal performance of fully digital beamforming. However, these methods require the perfect channel state information (CSI) to be acquired by the BS, which is very challenging in practice, especially when the number of RF chains is limited~\cite{roh2014millimeter}. By contrast, the codebook-based beamforming can obtain the optimal pair of analog precoder and analog combiner by searching the pre-defined codebook without knowing the exact channel. The most intuitive and optimal scheme is full search (FS) beamforming~\cite{kim2013tens}. However, its complexity increases exponentially with the number of RF chains and quantified bits of the angles of arrival and departure (AoA/AoDs). To reduce the searching complexity of codebook-based beamforming, some low-complexity schemes, such as the ones adopted by standards IEEE 802.15.3c~\cite{wang2009beam} and IEEE 802.11ad~\cite{cordeiro2010ieee}, have already been proposed. Furthermore, a multi-level codebook together with a ping-pong searching scheme is also proposed in~\cite{hur2013millimeter}. These schemes can reduce the searching complexity without obvious performance loss.  However, they usually involve a large number of iterations to exchange the information between the user and the BS, leading to a high overhead  for practical systems.

To reduce both the searching complexity and the overhead of codebook-based beamforming, in this paper, we propose a Turbo-like beamforming scheme based on tabu search algorithm~\cite{glover1989tabu} (called as Turbo-TS beamforming) with near-optimal\footnote{Note that ``near-optimal" means achieving the performance close to that of the optimal FS beamforming.} performance for mm-Wave massive MIMO systems. Specifically, the proposed Turbo-TS beamforming scheme is composed of the following two key components: 1) Based on the iterative information exchange between the BS and the user, we design a Turbo-like joint search scheme to find out the near-optimal pair of analog precoder and analog combiner; 2)  Inspired by TS algorithm in artificial intelligence, we develop a TS-based precoding/combining  to intelligently search the best precoder/combiner in each iteration of Turbo-like joint search with low complexity. Furthermore, the contributions of the proposed TS-based precoding/combing can be summarized in the following three aspects: 1)  Provide the appropriate definitions of neighborhood, cost, and stopping criterion involved in TS-based precoding/combing; 2) Take the exact solution instead of the conventional ``move" as tabu to guarantee a wider searching range; 3) Propose a restart method by selecting several different initial solutions uniformly distributed in the codebooks to further improve the performance. It is shown that the proposed Turbo-TS beamforming can considerably reduce the searching complexity. We verify through simulations that Turbo-TS beamforming can approach the performance of FS beamforming~\cite{kim2013tens}.

The rest of this paper is organized as follows. Section~\ref{S2} briefly introduces the system model of mmWave massive MIMO. Section~\ref{S3} specifies the proposed Turbo-TS beamforming. The simulation results of achievable rate are shown in Section~\ref{S4}. Finally, conclusions are drawn in Section~\ref{S5}.

{\it Notation}: Lower-case and upper-case boldface letters denote vectors and matrices, respectively;  ${( \cdot )^T}$, ${( \cdot )^H}$, ${( \cdot )^{ - 1}}$, and ${\det ( \cdot )}$  denote the transpose, conjugate transpose, inversion, and determinant of a matrix, respectively; ${\mathbb{E}( \cdot )}$  denotes the expectation; Finally, ${{\bf{I}}_N}$ is the  $ N \times N $  identity matrix.

\section{System Model}\label{S2}
We consider the mmWave massive MIMO system with beamforming as shown in Fig. 1, where the BS employs ${N_t}$ antennas and ${N_t^{{\rm{RF}}}}$ RF chains to simultaneously transmit ${N_s}$ data streams to a user with ${N_r}$ antennas and ${N_r^{{\rm{RF}}}}$ RF chains. To fully achieve the spatial multiplexing gain, we usually have ${N_t^{{\rm{RF}}} = N_r^{{\rm{RF}}} = {N_s}}$~\cite{el2013spatially}.
The ${N_s}$ independent transmitted data streams in the baseband firstly pass through ${N_t^{{\rm{RF}}}}$  RF chain to be converted into analog signals. After that, the output signals will be precoded by an ${{N_t} \times N_t^{{\rm{RF}}}}$ analog precoder ${{{\bf{P}}_{\rm A}}}$  as ${{\bf{x}} = {{\bf{P}}_{\rm A}}{\bf{s}}}$ before transmission, where ${{\bf{s}}}$ is the ${N_s \times 1}$ transmitted signal vector subject to the normalized power ${\mathbb{E}\left({{\bf{s}}{{\bf{s}}^H}} \right) = \frac{1}{{{N_s}}}{{\bf{I}}_{{N_s}}}}$. Note that the analog precoder ${{{\bf{P}}_{\rm A}}}$ is usually realized by a PS network with low hardware complexity~\cite{roh2014millimeter}, which requires that all elements of ${{{\bf{P}}_{\rm A}}}$ should satisfy ${{\left| {p_{i,j}^A} \right|^2} = \frac{1}{{{N_t}}}}$.
Under the narrowband block-fading massive MIMO channel~\cite{el2013spatially}, the ${N_r \times 1}$ received signal vector ${{\bf{r}}}$ at the user can be presented as
\begin{equation}\label{eq1}
{\bf{r}} = \sqrt \rho  {\bf{H}}{{\bf{P}}_{\rm A}}{\bf{s}} + {\bf{n}},
\end{equation}
where ${\rho }$ is the transmitted power, ${{\bf{H}} \in \mathbb{C}{^{N_r \times N_t}}}$ denotes the channel matrix which will be discussed in detail later in this section, and ${{\bf{n}} = {[{n_1}, \cdot  \cdot  \cdot ,{n_{N_r}}]^T}}$  is the additive white Gaussian noise (AWGN) vector, whose entries follow the independent and identical distribution (i.i.d.) ${{\cal C}{\cal N}(0,{\sigma ^2}{{\bf{I}}_{{N_r}}})}$.

At the user side, an ${{N_r} \times N_r^{{\rm{RF}}}}$ analog combiner ${{{\bf{C}}_{\rm A}}}$ is employed to process the received signal vector ${{\bf{r}}}$ as
\begin{equation}\label{eq2}
{\bf{y}} = {\bf{C}}_{\rm{A}}^H{\bf{r}} = \sqrt \rho  {\bf{C}}_{\rm{A}}^H{\bf{H}}{{\bf{P}}_{\rm{A}}}{\bf{s}} + {\bf{C}}_{\rm{A}}^H{\bf{n}},
\end{equation}
where the elements of ${{{\bf{C}}_{\rm A}}}$ have the similar constraints as that of ${{{\bf{P}}_{\rm A}}}$, i.e., ${{\left| {c_{i,j}^{\rm A}} \right|^2} = \frac{1}{{{N_r}}}}$.

Due to the limited number of significant scatters and serious antenna correlation of mmWave communication~\cite{pi2011introduction}, in this paper  we adopt the widely used geometric Saleh-Valenzuela channel model~\cite{el2013spatially}, where the channel matrix ${{\bf{H}}}$ can be presented as
\begin{equation}\label{eq3}
{\bf{H}} = \sqrt {\frac{{{N_t}{N_r}}}{L}} \sum\limits_{l = 1}^L {{\alpha _l}{{\bf{f}}_r}\left( {\phi _l^r} \right){\bf{f}}_t^H\left( {\phi _l^t} \right)},
\end{equation}
where ${L}$ is the number of significant scatters, and we usually have ${L \le \min \left( {{N_t},{N_r}} \right)}$ for mmWave communication systems due to the sparse nature of scatters, ${{\alpha _l} \in \mathbb{C}}$ is the gain of the  ${l}$th path including the path loss, ${\phi _l^t}$ and ${\phi _l^r}$ are the azimuth of AoDs/AoAs of the ${l}$th path, respectively. Finally, ${{{\bf{f}}_t}\left( {\phi _l^t} \right)}$  and  ${{{\bf{f}}_r}\left( {\phi _l^r} \right)}$ are the antenna array response vectors which depend on the antenna array structure at the BS and the user. When the widely used uniform linear arrays (ULAs) are considered, we have~\cite{el2013spatially}
\begin{equation}\label{eq4}
{{\bf{f}}_t}\left( {\phi _l^t} \right) = \frac{1}{{\sqrt {{N_t}} }}{\left[ {1,{e^{jkd\sin \left( {\phi _l^t} \right)}}, \cdot  \cdot  \cdot ,{e^{j\left( {{N_t} - 1} \right)kd\sin \left( {\phi _l^t} \right)}}} \right]^T},
\end{equation}
\begin{equation}\label{eq5}
{{\bf{f}}_r}\left( {\phi _l^r} \right) = \frac{1}{{\sqrt {{N_r}} }}{\left[ {1,{e^{jkd\sin \left( {\phi _l^r} \right)}}, \cdot  \cdot  \cdot ,{e^{j\left( {{N_r} - 1} \right)kd\sin \left( {\phi _l^r} \right)}}} \right]^T},
\end{equation}
where ${k = \frac{{2\pi }}{\lambda }}$, ${\lambda }$ denotes the wavelength of the signal, and ${d}$ is the antenna spacing.

\section{Near-Optimal Turbo-TS Beamforming With Low Complexity}\label{S3}
In this section, we first give a brief introduction of the codebook-based beamforming, which is widely used in mmWave massive MIMO systems. After that, a low-complexity near-optimal Turbo-TS beamforming scheme is proposed, which consists of Turbo-like joint search scheme and TS-based precoding/combining. Finally, the complexity analysis is provided to show the advantage of the proposed Turbo-TS beamforming scheme.

\subsection{Codebook-based beamforming}\label{S3.1}
According to the special characteristic of mmWave channel, the beamsteering codebook~\cite{kim2013tens} is widely used. Specifically, let ${{\cal F}}$ and ${{\cal W}}$ denote the beamsteering codebooks for the analog precoder and analog combiner, respectively. If we use ${B_t^{{\rm{RF}}}}$ (${B_r^{{\rm{RF}}}}$) bits to quantify the AoD (AoA), ${{\cal F}}$ (${{\cal W}}$) will consist of all the possible analog precoder (combiner) matrices ${{{\bf{P}}_{\rm A}}}$ (${{{\bf{C}}_{\rm A}}}$), which can be presented as~\cite{kim2013tens}
\begin{equation}\label{eq6}
{{\bf{P}}_{\rm A}} = \left[ {{{\bf{f}}_t}\left( {\bar \phi _1^t} \right),{{\bf{f}}_t}\left( {\bar \phi _2^t} \right), \cdot  \cdot  \cdot ,{{\bf{f}}_t}\left( {\bar \phi _{N_t^{{\rm{RF}}}}^t} \right)} \right],
\end{equation}
\begin{equation}\label{eq7}
{{\bf{C}}_{\rm A}} = \left[ {{{\bf{f}}_r}\left( {\bar \phi _1^r} \right),{{\bf{f}}_r}\left( {\bar \phi _2^r} \right), \cdot  \cdot  \cdot ,{{\bf{f}}_r}\left( {\bar \phi _{N_r^{{\rm{RF}}}}^r} \right)} \right],
\end{equation}
where the quantified AoD ${\bar \phi _i^t}$ for ${i = 1, \cdot  \cdot  \cdot ,N_t^{{\rm{RF}}}}$ at the BS has ${{{2^{B_t^{{\rm{RF}}}}}}}$ possible candidates, i.e., ${\bar \phi _i^t = \frac{{2\pi n}}{{{2^{B_t^{{\rm{RF}}}}}}}}$ where ${n \in \left\{ {1, \cdots {2^{B_t^{{\rm{RF}}}}}} \right\}}$. Similarly, the quantified AoA  ${\bar \phi _j^r}$ for ${j = 1, \cdot  \cdot  \cdot ,N_r^{{\rm{RF}}}}$ at the user has ${{{2^{B_r^{{\rm{RF}}}}}}}$ possible candidates, i.e.,  ${\bar \phi _j^r = \frac{{2\pi n}}{{{2^{B_r^{{\rm{RF}}}}}}}}$ where ${n \in \left\{ {1, \cdots {2^{B_r^{{\rm{RF}}}}}} \right\}}$. Thus, the cardinalities ${\left| {\cal F} \right|}$ of ${{\cal F}}$ and ${\left| {\cal W} \right|}$ of ${{\cal W}}$ are ${{2^{B_t^{{\rm{RF}}} \cdot N_t^{{\rm{RF}}}}}}$  and ${{2^{B_r^{{\rm{RF}}} \cdot N_r^{{\rm{RF}}}}}}$, respectively.  Then, by jointly searching ${{\cal F}}$ and ${{\cal W}}$, the optimal pair of analog precoder and analog combiner can be selected by maximizing the achievable rate as~\cite{el2013spatially}
\begin{equation}\label{eq8}
R = \mathop {\max }\limits_{{{\bf{P}}_{\rm{A}}} \in F,{{\bf{C}}_{\rm{A}}} \in W} {\log _2}\left( {\left| {{{\bf{I}}_{{N_s}}} + \frac{\rho }{{{N_s}}}{\bf{R}}_n^{ - 1}{\bf{C}}_{\rm{A}}^H{\bf{H}}{{\bf{P}}_{\rm{A}}}{\bf{P}}_{\rm{A}}^H{{\bf{H}}^H}{{\bf{C}}_{\rm{A}}}} \right|} \right) = \mathop {\max }\limits_{{{\bf{P}}_{\rm{A}}} \in F,{{\bf{C}}_{\rm{A}}} \in W} {\log _2}\left( {\varphi \left( {{{\bf{P}}_{\rm{A}}},{{\bf{C}}_{\rm{A}}}} \right)} \right),
\end{equation}
where ${{{\bf{R}}_n} = {\sigma ^2}{\bf{C}}_{\rm A}^H{{\bf{C}}_{\rm A}}}$ presents the covariance matrix of noise after combining, and
\begin{equation}\label{eq9}
\varphi \left( {{{\bf{P}}_{\rm A}},{{\bf{C}}_{\rm A}}} \right)\! =\! \left| {{{\bf{I}}_{{N_s}}}\! +\! \frac{\rho }{{{N_s}}}{\bf{R}}_n^{ - 1}{\bf{C}}_{\rm A}^H{\bf{H}}{{\bf{P}}_{\rm A}}{\bf{P}}_{\rm A}^H{{\bf{H}}^H}{{\bf{C}}_{\rm A}}} \right|
\end{equation}
is defined as the cost function. We can observe that to obtain the optimal pair of analog precoder and analog combiner, we need to exhaustively search the codebooks ${{\cal F}}$ and ${{\cal W}}$. When ${N_r^{{\rm{RF}}} = N_t^{{\rm{RF}}} = 2}$, ${B_t^{{\rm{RF}}} = B_r^{{\rm{RF}}} = 6}$, the totally required times of search is ${1.6 \times {10^7}}$, which is almost impossible in practice.
In this paper, we propose a Turbo-TS beamforming to reduce the searching complexity. The proposed Turbo-TS beamforming is composed of two key components, i.e., Turbo-like joint search scheme and TS-based precoding/combining, which will be described in detail in the following Section III-B and Section III-C, respectively.

\subsection{Turbo-like joint search scheme}\label{S3.1}
Based on the idea of the information interaction in the well-known turbo equalizer, we propose a Turbo-like joint search scheme to find out the near-optimal pair of analog precoder and analog combiner, which is shown in Fig. 2. Let ${{\bf{P}}_{\rm A}^{{\rm{opt}},k}}$ and ${{\bf{C}}_{\rm A}^{{\rm{opt}},k}}$ denote the near-optimal analog precoder and analog combiner obtained in the ${k}$th iteration, respectively, where ${k = 1, 2, \cdots, K}$, and ${K}$ is the pre-defined maximum number of iterations. Firstly, the BS selects an initial precoder ${{\bf{P}}_{\rm A}^{{\rm{opt}},0}}$, which can be an arbitrary candidate in ${{\cal F}}$, to transmit a training sequence to the user. Then the user can search the best analog combiner ${{\bf{C}}_{\rm A}^{{\rm{opt}},1}}$. After that, the user uses ${{\bf{C}}_{\rm A}^{{\rm{opt}},1}}$ to transmit a training sequence to the BS, and in return the BS can search the best analog precoder ${{\bf{P}}_{\rm A}^{{\rm{opt}},1}}$.  We repeat such iteration for ${K}$ times in a similar way as the turbo equalizer, and output ${{\bf{P}}_{\rm A}^{{\rm{opt}},K}}$ and ${{\bf{C}}_{\rm A}^{{\rm{opt}},K}}$ as the final pair of analog precoder and analog combiner, which is expected to achieve the near-optimal performance as will be verified later in Section~\ref{S4}. Note that in each iteration, searching the best analog precoder (combiner)  after a potential analog combiner (precoder) has been selected from the codebook ${\cal W}$ (${\cal F}$) can be realized  by the proposed TS-based analog precoding/combinging with low complexity, which will be described in detail in the next subsection.

\subsection{TS-based precoding/combining}\label{S3.2}
In this subsection, we first focus on the process of searching the best analog precoder ${{{\bf{P}}_{\rm A}}}$  after a potential analog combiner ${{{\bf{C}}_{\rm A}}}$ has been selected. The process of searching the best analog combiner ${{{\bf{C}}_{\rm A}}}$ after a certain analog precoder ${{{\bf{P}}_{\rm A}}}$ has been selected can be derived in the similar way.


The basic idea of the proposed TS-based analog precoding can be described as follows. TS-based analog precoding  starts from an initial solution, i.e., an analog precoder matrix selected from the codebook ${{\cal F}}$, and defines a neighborhood around it (several analog precoder matrices from ${{\cal F}}$ based on a neighboring criterion). After that, it selects the most appropriate solution among the neighborhood as the starting point for the next iteration, even if it is not the global optimum. During the search in the neighborhood, TS attempts to escape from the local optimum by utilizing the concept of ``tabu", whose definition can be changed according to different criterions (e.g., convergence speed, complexity, etc). This process will be continued until a certain stopping criterion is satisfied, and finally the best solution among all iterations will be declared as the final solution. Next, five important aspects of the proposed TS-based precoding, including neighborhood definition, cost computation, tabu, stopping criterion, and TS algorithm, will be explained in detail as follows.

\emph{1) Neighborhood definition:}  Note that the ${m}$th column of analog precoder ${{{\bf{P}}_{\rm A}}}$ can be presented by an index ${q_m \in \left\{ {1,2, \cdot  \cdot  \cdot ,{2^{B_t^{{\rm{RF}}}}}} \right\}}$, which corresponds to the vector ${{{\bf{f}}_t}\left( {\frac{{2\pi q_m}}{{{2^{B_t^{{\rm{RF}}}}}}}} \right)}$ as defined in (\ref{eq4}) and (\ref{eq6}). Then an analog precoder is defined as a neighbor of ${{{\bf{P}}_{\rm A}}}$ if: i) it has only one column that is different from the corresponding column in ${{{\bf{P}}_{\rm A}}}$; ii) the index difference between the two corresponding columns equals one. For example, when ${N_t^{{\rm{RF}}} = 2}$  and ${B_t^{{\rm{RF}}} = 3}$, for a possible analog precoder ${{{\bf{P}}_{\rm A}} = \left[ {{{\bf{f}}_t}\left( {\frac{{3\pi }}{4}} \right),{{\bf{f}}_t}\left( {\frac{{7\pi }}{4}} \right)} \right]}$, another precoder ${\left[ {{{\bf{f}}_t}\left( {\frac{{2\pi }}{4}} \right),{{\bf{f}}_t}\left( {\frac{{7\pi }}{4}} \right)} \right]}$ is a neighbor of ${{{\bf{P}}_{\rm A}}}$.

Let ${{{\bf{P}}_{\rm A}^{\left( i \right)}}}$ denote the starting point in the ${i}$th iteration of the proposed TS-based analog precoding, and ${{\cal V}\left( {{\bf{P}}_{\rm A}^{\left( i \right)}} \right)\! =\! \left\{ {{\bf{V}}_1^{\left( i \right)},{\bf{V}}_2^{\left( i \right)}, \cdot  \cdot  \cdot {\bf{V}}_{\left| {\cal V} \right|}^{\left( i \right)}} \right\}}$ presents the neighborhood of ${{{\bf{P}}_{\rm A}^{\left( i \right)}}}$, where ${{\left| {\cal V} \right|}}$ is the cardinality of ${{\cal V}}$. According to the neighborhood definition above, it is obvious that ${\left| {\cal V} \right| = 2N_t^{{\rm{RF}}}}$. We then define that the ${u}$th neighbor  in ${{\cal V}\left( {{\bf{P}}_{\rm A}^{\left( i \right)}} \right)}$ is different from ${{{\bf{P}}_{\rm A}^{\left( i \right)}}}$ in the ${\left\lceil {\frac{u}{2}} \right\rceil }$th column, and the index of the corresponding column is ${{q_{\left\lceil {u/2} \right\rceil }} + {\left( { - 1} \right)^{\bmod \left( {u,2} \right)}}}$, where ${{q_{\left\lceil {u/2} \right\rceil }}}$ is the index of this column. To avoid overflow of the definition above, we set
\begin{equation}\label{eq11}
1 + {\left( { - 1} \right)^{\bmod \left( {u,2} \right)}} = \max \left( {1 + {{\left( { - 1} \right)}^{\bmod \left( {u,2} \right)}},1} \right),
\end{equation}
\begin{equation}\label{eq12}
{2^{B_t^{{\rm{RF}}}}} + {\left( { - 1} \right)^{\bmod \left( {u,2} \right)}} = \min \left( {{2^{B_t^{{\rm{RF}}}}} + {{\left( { - 1} \right)}^{\bmod \left( {u,2} \right)}},{2^{B_t^{{\rm{RF}}}}}} \right).
\end{equation}
For example, the neighborhood of one analog precoder ${{{\bf{P}}_{\rm A}^{\left( i \right)}}\! =\! \left[ {{{\bf{f}}_t}\left( {\frac{{3\pi }}{4}} \right),{{\bf{f}}_t}\left( {\frac{{7\pi }}{4}} \right)} \right]}$ is  ${{{{\bf{V}}_1^{\left( i \right)}}}\! =\! \left[ {{{\bf{f}}_t}\left( {\frac{{2\pi }}{4}} \right),{{\bf{f}}_t}\left( {\frac{{7\pi }}{4}} \right)} \right]}$, ${{{{\bf{V}}_2^{\left( i \right)}}}\! =\! \left[ {{{\bf{f}}_t}\left( {\frac{{4\pi }}{4}} \right),{{\bf{f}}_t}\left( {\frac{{7\pi }}{4}} \right)} \right]}$, ${{{{\bf{V}}_3^{\left( i \right)}}}\! =\! \left[ {{{\bf{f}}_t}\left( {\frac{{3\pi }}{4}} \right),{{\bf{f}}_t}\left( {\frac{{6\pi }}{4}} \right)} \right]}$, and ${{{{\bf{V}}_4^{\left( i \right)}}}\! =\! \left[ {{{\bf{f}}_t}\left( {\frac{{3\pi }}{4}} \right),{{\bf{f}}_t}\left( {\frac{{8\pi }}{4}} \right)} \right]}$.

\emph{2) Cost computation:} We define the value of the cost function ${\varphi \left( {{{\bf{P}}_{\rm{A}}},{{\bf{C}}_{\rm{A}}}} \right)}$ in (\ref{eq9}) as the reliability metric of a possible solution, i.e., a solution ${{{\bf{P}}_{\rm{A}}}}$ leading to a larger value of ${\varphi \left( {{{\bf{P}}_{\rm{A}}},{{\bf{C}}_{\rm{A}}}} \right)}$ is a better solution. Further, according to the neighborhood definition, we can observe that once we obtain the cost of ${{{\bf{P}}_{\rm A}}}$, we do not need to recompute (\ref{eq9}) to obtain the cost of its neighborhood through information exchange between the BS and the user. This is due to the fact that the neighbor ${{{\bf{V}}_u}}$ of ${{{\bf{P}}_{\rm A}}}$ only has the ${\left\lceil {\frac{u}{2}} \right\rceil }$th column that is different from the corresponding one in ${{{\bf{P}}_{\rm A}}}$, then the updated effective channel matrix ${{\bf{C}}_{\rm A}^H{\bf{H}}{{\bf{V}}_u}}$ in (\ref{eq9}) also has the ${\left\lceil {\frac{u}{2}} \right\rceil }$th column that is different from the corresponding one in the original effective channel matrix ${{\bf{C}}_{\rm A}^H{\bf{H}}{{\bf{P}}_{\rm A}}}$, where such difference can be easily calculated since ${{{\bf{P}}_{\rm A}}}$ and ${{{\bf{V}}_u}}$ are known. More importantly, this special property indicates that for the proposed TS-based analog precoding, we can only estimate the effective channel matrix ${{\bf{C}}_{\rm A}^H{\bf{H}}{{\bf{P}}_{\rm A}}}$ of size ${N_r^{{\rm{RF}}} \times N_t^{{\rm{RF}}}}$ through
time-domain and/or frequency-domain training sequence~\cite{dai13}, whose dimension is much lower than the original dimension ${N_r \times N_t}$ of the exact channel matrix ${{\bf{H}}}$.

\emph{3) Tabu:} In the conventional TS algorithm~\cite{glover1989tabu}, the tabu is usually defined as the ``move", which can be regarded as the direction from one solution to another one for the analog precoding problem. The ``move" can be denoted by ${\left( {a,b} \right)}$, where ${a = 1, \cdot  \cdot  \cdot ,N_t^{{\rm{RF}}}}$ denotes that the ${a}$th column of the original solution is different from that of the current solution, ${b \in \left\{ { - 1,1} \right\}}$ means the changed index of this particular column  from the original solution to the current solution. Consider the example above, the ``move" (direction) from  ${\left[ {{{\bf{f}}_t}\left( {\frac{{3\pi }}{4}} \right),{{\bf{f}}_t}\left( {\frac{{7\pi }}{4}} \right)} \right]}$ to ${\left[ {{{\bf{f}}_t}\left( {\frac{{2\pi }}{4}} \right),{{\bf{f}}_t}\left( {\frac{{7\pi }}{4}} \right)} \right]}$ can be written as ${\left( {1, - 1} \right)}$. Regarding the ``move" as tabu can save storage of the tabu list, since it only requires a tabu list ${{\bf{t}}}$ of size ${2N_t^{{\rm{RF}}} \times 1}$, whose element takes the value from ${\left\{ {0,1} \right\}}$ to indicate whether a move is tabu or not (i.e., 1 is tabu, and 0 is unconstrained). However, as shown in Fig. 3 (a), this method may lead to the unexpected fact that one solution will be searched twice, and the cost function of the same neighborhood will be computed again.
To solve this problem, we propose to take the exact solution as tabu. Specifically, let ${p = 1,2, \cdot  \cdot  \cdot ,{2^{B_t^{{\rm{RF}}} \cdot N_t^{{\rm{RF}}}}}}$ present the index of a candidate of the analog precoder (solution) out of ${{\cal F}}$ with ${{2^{B_t^{{\rm{RF}}} \cdot N_t^{{\rm{RF}}}}}}$ possible candidates. Particularly, ${p}$ can be calculated by each column index ${q_m}$ (${1 \le {q_m} \le N_t^{{\rm{RF}}}}$) \footnote{It is worth pointing out that to fully achieve the spatial multiplexing gain, the column index ${q_m}$ should be different for different RF chains, i.e., ${{q_1} \ne {q_2} \ne  \cdots  \ne {q_{N_t^{{\rm{RF}}}}}}$. All the possible precoder/combiner matrices that do not obey this constraint will be declared as ``tabu" to avoid being searched.} of this analog precoder as
\begin{equation}\label{eq13}
p = \sum\limits_{m = 1}^{N_t^{{\rm{RF}}}} {\left( {{q_m} - 1} \right){{\left( {{2^{B_t^{{\rm{RF}}}}}} \right)}^{N_t^{{\rm{RF}}} - m}}}  + 1.
\end{equation}
For example, when ${B_t^{{\rm{RF}}} = 3}$ and ${N_t^{{\rm{RF}}} = 2}$, if an analog precoder has the column indexes ${\left\{ {2,7} \right\}}$, then the index of this analog precoder in ${{\cal F}}$ is ${p = 15}$ according to (\ref{eq13}). In this way, our method can efficiently avoid one solution being searched twice, and therefore a wider searching range can be achieved as shown in Fig. 3 (b). Note that the only cost of our method is the increased storage size of the tabu list ${{\bf{t}}}$ from ${2N_t^{{\rm{RF}}}}$ to ${{2^{B_t^{{\rm{RF}}} \cdot N_t^{{\rm{RF}}}}}}$.

\emph{4) Stopping criterion:} We define ${{\rm{flag}}}$ as a parameter to indicate how long (in terms of number of iterations) the global optimal solution has not been updated. That means in the current iteration, if a suboptimal solution is selected as the starting point for the next iteration,  we have ${\rm{flag} = \rm{flag} + 1}$, otherwise, if the global optimal solution is selected, we set ${\rm{flag} = 0}$. Based on this mechanism, TS-based analog precoding will be terminated when either of the following two conditions is satisfied: i) The total number of iterations reaches the pre-defined maximum number of iterations ${{\rm{max\_iter}}}$; ii) The number of iterations for the global optimal solution not being updated reaches the pre-defined maximum value ${{\rm{max\_len}}}$, i.e., ${{\rm{flag}} = {\rm{max\_len}}}$. Note that we usually set ${{\rm{max\_len}} < {\rm{max\_iter}}}$, which means if TS-based analog precoding has already found the optimal solution at the beginning, all the starting points in following iterations will be suboptimal, so we don't need to wait ${{\rm{max\_iter}}}$ iterations. Therefore, the average searching complexity can be reduced further.

\emph{5) Tabu search algorithm:} Let ${{{\bf{G}}^{(i)}}}$ denote the analog precoder achieving the maximum cost function (\ref{eq9}) that has been found until the ${i}$th iteration. TS-based analog precoding starts with the initial solution ${{\bf{P}}_{\rm A}^{(0)}}$. Note that in order to improve the performance of TS-based analog precoding, we can select ${M}$ different initial solutions uniformly distributed in ${{\cal F}}$ to start TS-based analog precoding ${M}$ times, then, the best one out of ${M}$ obtained solutions will be declared as the final analog precoder. For each initial solution, we set ${{{\bf{G}}^{(0)}} = {\bf{P}}_{\rm A}^{(0)}}$, ${{\rm{flag}} = 0}$. Besides, all the elements of the tabu list ${{\bf{t}}}$ are set as zero. Considering the ${i}$th iteration, TS-based analog precoding executes as follows:

\emph{Step 1:} Compute the cost function (\ref{eq9}) of the ${2N_t^{{\rm{RF}}}}$ neighbors of ${{\bf{P}}_{\rm A}^{(i)}}$ given the effective channel matrix ${{\bf{C}}_{\rm A}^H{\bf{H}}{\bf{P}}_{\rm A}^{(i)}}$. Let
\begin{equation}\label{eq14}
{{\bf{V}}^1} = \arg \mathop {\max }\limits_{1 \le u \le 2N_t^{{\rm{RF}}}} \varphi \left( {{{\bf{V}}_u},{{\bf{C}}_{\rm{A}}}} \right).
\end{equation}
Calculate the index ${{p^1}}$ of ${{{\bf{V}}^1}}$ in ${{\cal F}}$ according to (\ref{eq13}). Then, ${{{\bf{V}}^1}}$ will be selected as the starting point for the next iteration when either of the following two conditions is satisfied:
\begin{equation}\label{eq15}
\varphi \left( {{{\bf{V}}^1},{{\bf{C}}_{\rm{A}}}} \right) > \varphi \left( {{{\bf{G}}^{\left( i \right)}},{{\bf{C}}_{\rm{A}}}} \right),
\end{equation}
\begin{equation}\label{eq16}
{\bf{t}}\left( {{p^1}} \right) = 0.
\end{equation}
If ${{{{\bf{V}}^1}}}$ cannot be selected, we find the second best solution as
\begin{equation}\label{eq17}
{{\bf{V}}^2} = \arg \mathop {\max }\limits_{\mathop {1 \le u \le 2N_t^{{\rm{RF}}}}\limits_{{{\bf{V}}_u} \ne {{\bf{V}}^1}} } \varphi \left( {{{\bf{V}}_u},{{\bf{C}}_{\rm{A}}}} \right).
\end{equation}
Then we decide whether ${{{\bf{V}}^2}}$ can be selected by checking (\ref{eq15}) and (\ref{eq16}). This procedure will be continued until one solution ${{\bf{V}}'}$ is selected as the starting point for the next iteration. Note that if there is no solution satisfying (\ref{eq15}) and (\ref{eq16}), all the corresponding elements of the tabu list ${{\bf{t}}}$ will be set to zero, and the same procedure above will be repeated.

\emph{Step 2:} After a solution has been selected as the starting point, i.e., ${ {\bf{P}}_{\rm A}^{\left( {i + 1} \right)} = {\bf{V}}' }$, we set
\begin{equation}\label{eq18}
\left\{ \begin{array}{l}
{\bf{t}}\left( {p'} \right) = 0,\;{{\bf{G}}^{\left( {i + 1} \right)}} = {\bf{P}}_{\rm{A}}^{\left( {i + 1} \right)},\quad {\rm if}\;\varphi \left( {{\bf{P}}_{\rm{A}}^{\left( {i + 1} \right)},{{\bf{C}}_{\rm{A}}}} \right) > \;\varphi \left( {{{\bf{G}}^{\left( i \right)}},{{\bf{C}}_{\rm{A}}}} \right),\\
{\bf{t}}\left( {p'} \right) = 1,\;{{\bf{G}}^{\left( {i + 1} \right)}} = {{\bf{G}}^{\left( i \right)}},\quad \;\;\;{\rm if}\;\varphi \left( {{\bf{P}}_{\rm{A}}^{\left( {i + 1} \right)},{{\bf{C}}_{\rm{A}}}} \right) \le \;\varphi \left( {{{\bf{G}}^{\left( i \right)}},{{\bf{C}}_{\rm{A}}}} \right).
\end{array} \right.
\end{equation}
TS-based analog precoding will be terminated in \emph{Step 2} and output ${{{\bf{G}}^{\left( {i + 1} \right)}}}$ as the final solution if the stopping criterion is satisfied. Otherwise it will go back to \emph{Step 1} and repeat the procedure above until it satisfies the stopping criterion.

It is worth pointing out that searching the near-optimal analog combiner ${{{\bf{C}}_{\rm A}}}$ after a certain analog precoder ${{{\bf{P}}_{\rm A}}}$ has been selected can be also solved by similar procedure described above, where the definitions such as neighborhood should be changed accordingly to search the near-optimal analog combiner ${{{\bf{C}}_{\rm A}}}$.


\subsection{Complexity analysis}\label{S3.3}
In this subsection, we provide the complexity comparison between the proposed Turbo-TS beamforming and the conventional FS beamforming. It is worth pointing out that although the proposed Turbo-TS beamforming requires some extra information exchange between the BS and the UE (${K}$ times of iterations) as discussed in Section III-B, the corresponding overhead is trivial compared with the searching complexity, since ${K}$ is usually small (e.g., ${K=4}$ as will be verified by simulation results). Therefore, in this section we evaluate the complexity as the total number of solutions need to be searched. It is obvious that the searching complexity of FS beamforming ${{C_{{\rm{FS}}}}}$ is
\begin{equation}\label{eq19}
{C_{{\rm{FS}}}} = \left( {\begin{array}{*{20}{c}}
{N_t^{{\rm{RF}}}}\\
{{2^{B_t^{{\rm{RF}}}}}}
\end{array}} \right) \times \left( {\begin{array}{*{20}{c}}
{N_r^{{\rm{RF}}}}\\
{{2^{B_r^{{\rm{RF}}}}}}
\end{array}} \right).
\end{equation}
By contrast, the searching complexity of the proposed Turbo-TS beamforming ${{C_{{\rm{TS}}}}}$ is
\begin{equation}\label{eq20}
{C_{{\rm{TS}}}} = \left( {2N_t^{{\rm{RF}}} \cdot {\rm{max\_iter}} + 2N_r^{{\rm{RF}}} \cdot {\rm{max\_iter}}} \right)MK.
\end{equation}

Comparing (\ref{eq19}) and (\ref{eq20}), we can observe that the complexity of Turbo-TS beamforming is linear with ${N_t^{{\rm{RF}}}}$ and ${N_r^{{\rm{RF}}}}$, and it is independent of ${{B_t^{{\rm{RF}}}}}$ and ${{B_r^{{\rm{RF}}}}}$, which indicates that Turbo-TS beamforming enjoys a much lower complexity than FS beamforming. Table I shows the comparison of the searching complexity between Turbo-TS beamforming and FS beamforming when the numbers of RF chains at the BS and the user are ${N_t^{{\rm{RF}}} = N_r^{{\rm{RF}}}=2}$, where three cases are considered: 1) For ${B_t^{{\rm{RF}}} = B_r^{{\rm{RF}}} = 4}$, we set ${{\rm{max\_iter}} = 500}$ and ${{\rm{max\_len}} = 100}$, and uniformly select ${M=1}$ different initial solutions to initiate the TS-based precoding/combining; 2) For ${B_t^{{\rm{RF}}} = B_r^{{\rm{RF}}} = 5}$, we set ${{\rm{max\_iter}} = 1000}$, ${{\rm{max\_len}} = 200}$, and ${M=2}$; 3) For ${B_t^{{\rm{RF}}} = B_r^{{\rm{RF}}} = 6}$, we set ${{\rm{max\_iter}} = 3000}$, ${{\rm{max\_len}} = 600}$, and ${M=5}$. Besides, for all these cases above, we set the total number of iterations ${K=4}$ for the Turbo-like joint search scheme.
From Table I, we can observe that the proposed Turbo-TS beamforming scheme has much lower searching complexity than the conventional FS beamforming, e.g., when ${B_t^{{\rm{RF}}}\! =\! B_r^{{\rm{RF}}}\! =\! 6}$, the searching complexity of Turbo-TS beamforming is only 2.1\% of that of FS beamforming.

\section{Simulation Results}\label{S4}
We evaluate the performance of the proposed Turbo-TS beamforming in terms of the achievable rate. Here we also provide the performance of the recently proposed beam steering scheme~\cite{el2012capacity} with continuous angles as the benchmark for comparison, since it can be regarded as the upper bound of the proposed Turbo-TS beamforming with quantified AoA/AoDs. The system parameters for simulation are described as follows: The carrier frequency is set as 28GHz; We generate the channel matrix according to the channel model~\cite{el2013spatially} described in Section~\ref{S2}; The AoAs/AoDs are assumed to follow the uniform distribution within ${[0, \pi]}$; The complex gain ${{\alpha _l}}$  of the ${l}$th path follows ${{\alpha _l} \sim {\cal C}{\cal N}\left( {0,1} \right)}$, and the total number of scattering propagation paths is set as ${L = 3}$; Both the transmit and receive antenna arrays are ULAs with antenna spacing ${d = \lambda /2}$. Three cases of quantified bits per AoAs/AoDs, i.e., ${B_t^{{\rm{RF}}} = B_r^{{\rm{RF}}} = 4}$, ${B_t^{{\rm{RF}}} = B_r^{{\rm{RF}}} = 5}$, and ${B_t^{{\rm{RF}}} = B_r^{{\rm{RF}}} = 6}$ are evaluated;  SNR is defined as ${\frac{\rho }{{{\sigma ^2}}}}$; Additionally, the parameters used for the proposed TS-based precoding/combing are the same as those in Section III-D.

At first, we provide the achievable rate performance of Turbo-TS beamforming against different parameters to explain why we choose these values as listed in Section III-D. Fig. 4 shows a example when ${{N_r} \times {N_t} = 16 \times 64}$, ${N_r^{{\rm{RF}}} = N_t^{{\rm{RF}}} = {N_s} = 2}$, ${B_t^{{\rm{RF}}} = B_r^{{\rm{RF}}} = 6}$, and SNR = 0 dB. We can observe that when ${{\rm{max\_iter}} = 3000}$ (Fig. 4 (a)), ${{\rm{max\_len}} = 600}$ (Fig. 4 (b)), ${M=5}$ (Fig. 4 (c)), and ${K=4}$ (Fig. 4 (d)), the proposed Turbo-TS beamforming can achieve more than 90\% of the rate of FS beamforming, which verifies the rationality of our selection.


Fig. 5 shows the achievable rate comparison between the conventional FS beamforming and the proposed Turbo-TS beamforming for an ${{N_r} \times {N_t} = 16 \times 64}$  mmWave massive MIMO system with ${N_r^{{\rm{RF}}} = N_t^{{\rm{RF}}} = {N_s} = 2}$. We can observe that Turbo-TS beamforming can approach the achievable rate of FS beamforming without obvious performance loss. For example, when ${B_t^{{\rm{RF}}} = B_r^{{\rm{RF}}} = 4}$ and SNR = 0 dB, the rate achieved by Turbo-TS beamforming is 7 bit/s/Hz, which is quite close to 7.2 bit/s/Hz achieved by FS beamforming. When the number of quantified bits per AoAs/AoDs increases, both Turbo-TS beamforming and FS beamforming can achieve better performance close to the beam steering scheme with continuous AoAs/AoDs~\cite{el2012capacity}. Meanwhile, Turbo-TS beamforming can still guarantee the satisfying performance quite close to FS beamforming. Considering the considerably reduced searching complexity of Turbo-TS beamforming, we can conclude that the proposed Turbo-TS beamforming achieves a much better trade-off between performance and complexity.

Fig. 6 shows the achievable rate comparison for an ${{N_r} \times {N_t} = 32 \times 128}$  mmWave massive MIMO system, where the number of RF chains is still set as ${N_r^{{\rm{RF}}} = N_t^{{\rm{RF}}} = {N_s} = 2}$. From Fig. 6, we can observe similar trends as those from Fig. 5. More importantly, comparing Fig. 5 and Fig. 6, we can find that the performance of the proposed Turbo-TS beamforming can be improved by increasing the number of low-cost antennas instead of increasing the number of
expensive RF chains. For example, when ${{N_r} \times {N_t} = 16 \times 64}$, ${B_t^{{\rm{RF}}} = B_r^{{\rm{RF}}} = 6}$, and SNR = 0 dB, Turbo-TS beamforming can achieve the rate of 10.1 bit/s/Hz , while when ${{N_r} \times {N_t} = 32 \times 128}$, the achievable rate can be increased to 14 bit/s/Hz without increasing the number of RF chains.

\vspace*{-3mm}
\section{Conclusions}\label{S5}
In this paper, we propose a Turbo-TS beamforming scheme, which consists of two key components:  1) a Turbo-like joint search scheme relying on the iterative information exchange between the BS and the user; 2) a TS-based precoding/combining utilizing the idea of local search to find the best precoder/combiner in each iteration of Turbo-like joint search with low complexity. Analysis has shown that the complexity of the proposed scheme is linear with ${N_t^{{\rm{RF}}}}$ and ${N_r^{{\rm{RF}}}}$, and it is independent of ${{B_t^{{\rm{RF}}}}}$ and ${{B_r^{{\rm{RF}}}}}$, which can considerably reduce the complexity of conventional schemes. Simulation results have verified that the near-optimal performance of the proposed Turbo-TS beamforming. Our further work will focus on extending the proposed Turbo-TS beamforming to the multi-user scenario.

\vspace*{-1mm}
\bibliography{IEEEabrv,Gao1Ref}

\newpage

\vbox{}
\vbox{}

\begin{figure*}[h]
\begin{center}
\hspace*{0mm}\includegraphics[width=1\linewidth]{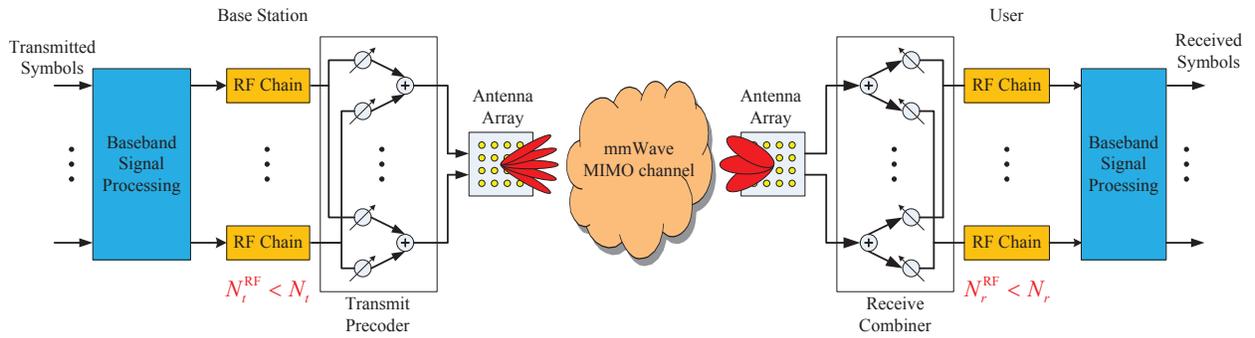}
\end{center}
\caption{Architecture of mmWave massive MIMO system with beamforming.} \label{FIG1}
\vspace*{+3mm}
\end{figure*}

\vbox{}
\vbox{}

\begin{figure}[h]
\begin{center}
\hspace*{0mm}\includegraphics[width=0.6\linewidth]{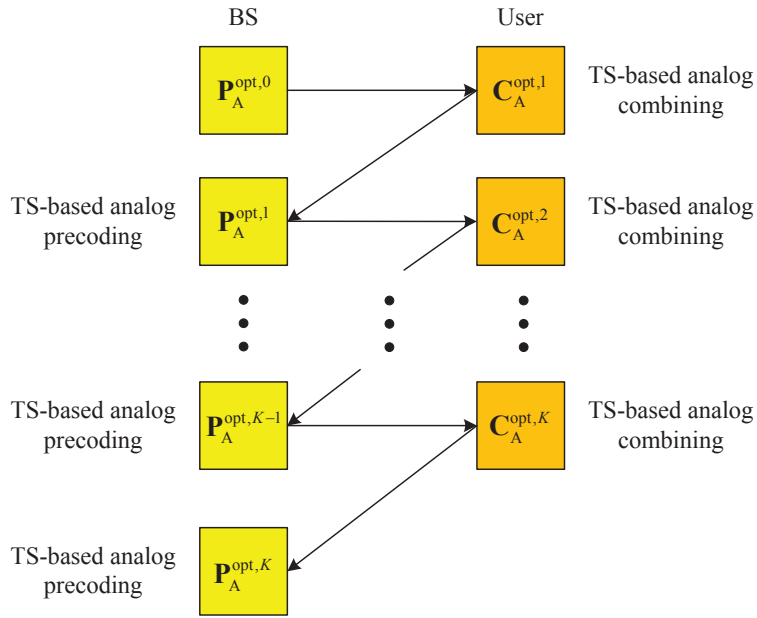}
\end{center}
\caption{Proposed Turbo-like joint search scheme.} \label{FIG4}
\end{figure}

\begin{figure}[tp]
\begin{center}
\hspace*{-6mm}\includegraphics[width=0.6\linewidth]{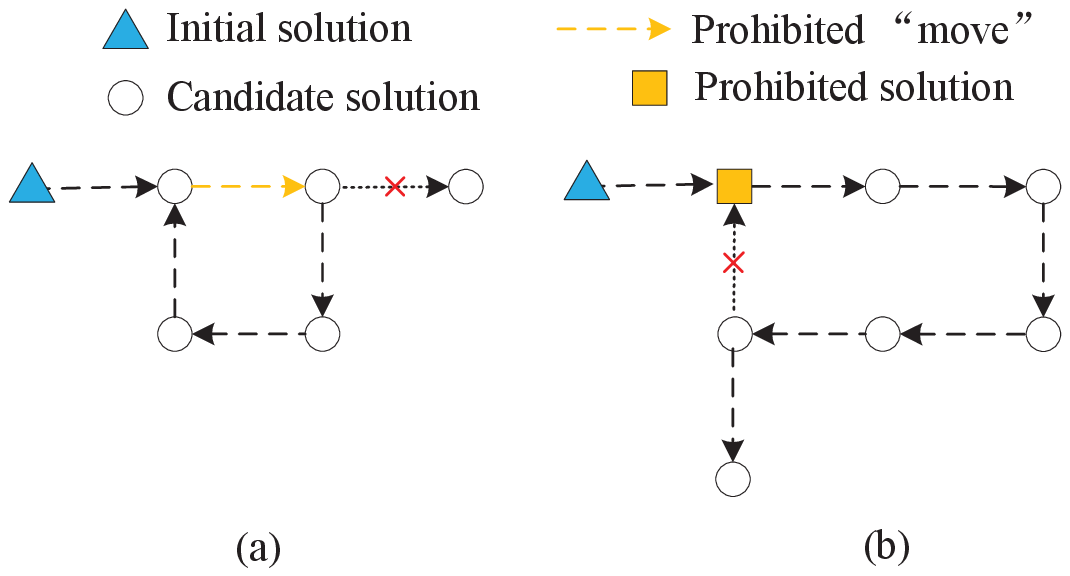}
\end{center}
\caption{Illustration of how the solution tabu can avoid one solution being searched twice; (a) Conventional ``move" tabu; (b) Proposed solution tabu.} \label{FIG3}
\vspace*{+3mm}
\end{figure}

\begin{table}[h]
\setlength{\abovecaptionskip}{-10pt}
\setlength{\belowcaptionskip}{0pt}
\caption{Complexity comparison} \label{TAB1}
\begin{center}
\begin{threeparttable}
\begin{tabular}{*{1}{L{3cm}}*{2}{C{3cm}}*{1}{C{3cm}}}
\toprule[1pt]
 & Conventional FS beamforming~\cite{kim2013tens} & Proposed Turbo-TS beamforming & Complexity ratio (TS/FS) \\
\hline \\ [-2 ex]
 ${B_t^{{\rm{RF}}} = B_r^{{\rm{RF}}} = 4}$ & 57600 & 16000 & 27.8 \%\\
 ${B_t^{{\rm{RF}}} = B_r^{{\rm{RF}}} = 5}$ & 984064 & 64000 & 6.5 \%\\
 ${B_t^{{\rm{RF}}} = B_r^{{\rm{RF}}} = 6}$ & 16257024 & 480000 & 2.9 \%\\
\toprule[1pt]
\end{tabular}
\end{threeparttable}
\end{center}
\end{table}

\begin{figure}[h]
\setlength{\abovecaptionskip}{-10pt}
\setlength{\belowcaptionskip}{0pt}
\begin{center}
\vspace*{0mm}\includegraphics[width=1\linewidth]{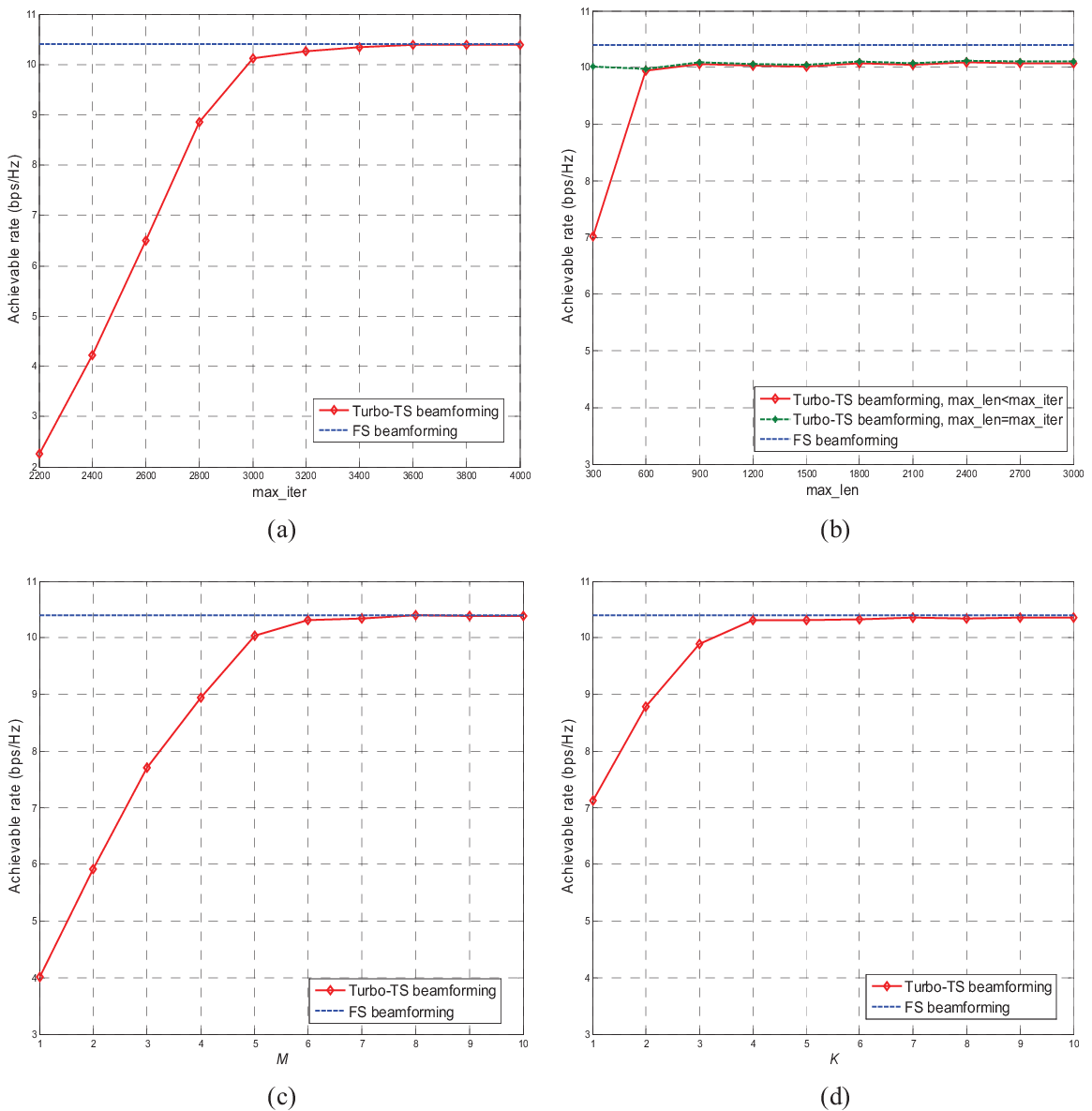}
\end{center}
\caption{Achievable rate of Turbo-TS beamforming against different parameters: (a) ${{\rm{max\_iter}}}$; (b) ${{\rm{max\_len}}}$; (c) ${M}$; (d) ${K}$.} \label{FIG4}
\vspace*{+3mm}
\end{figure}

\begin{figure}[h]
\setlength{\abovecaptionskip}{-10pt}
\setlength{\belowcaptionskip}{0pt}
\begin{center}
\vspace*{0mm}\includegraphics[width=0.6\linewidth]{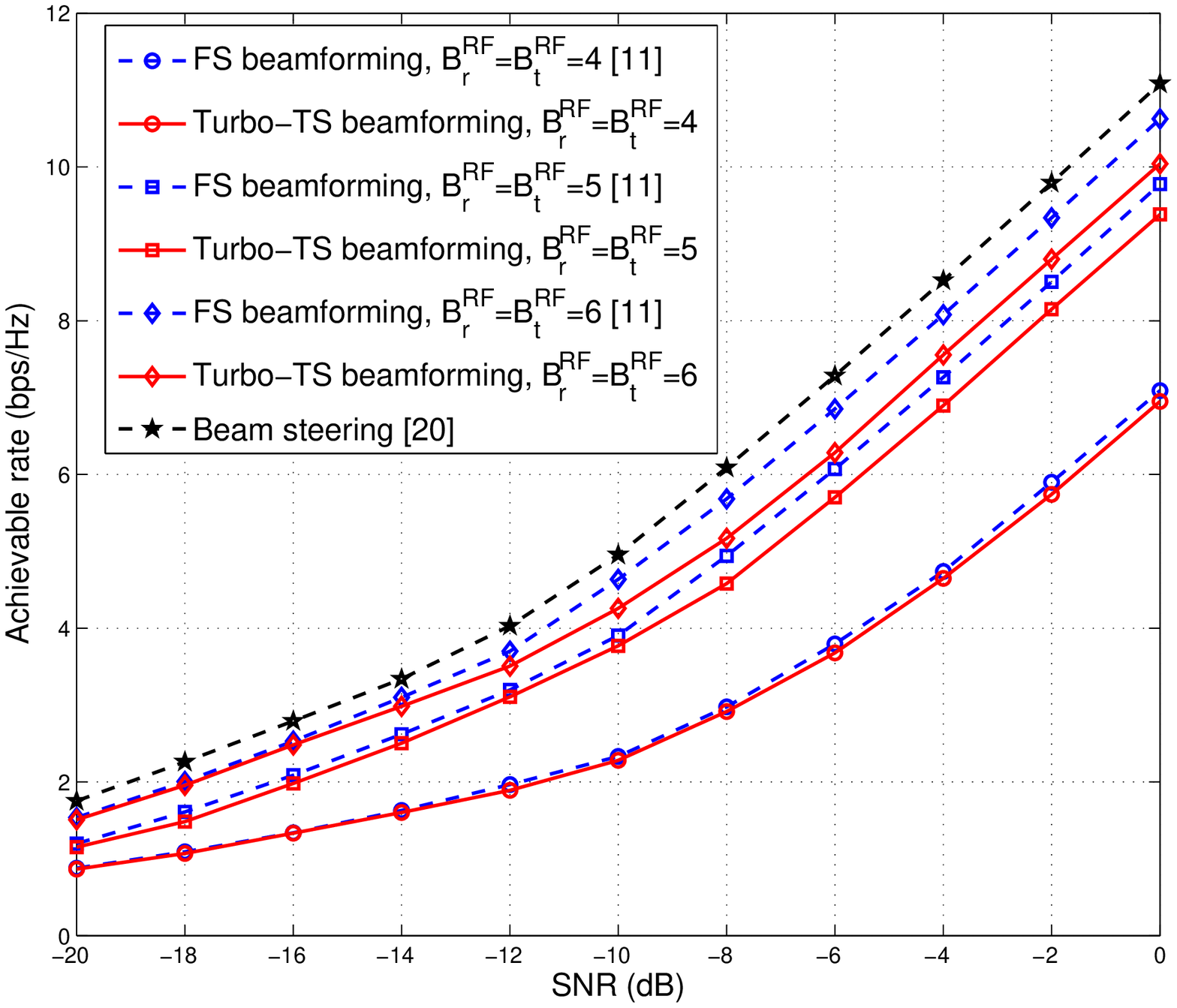}
\end{center}
\caption{Achievable rate comparison for an ${{N_r} \times {N_t} = 16 \times 64}$ mmWave massive MIMO system with ${N_r^{{\rm{RF}}} = N_t^{{\rm{RF}}} = {N_s} = 2}$.} \label{FIG4}
\vspace*{+3mm}
\end{figure}

\begin{figure}[h]
\setlength{\abovecaptionskip}{-10pt}
\setlength{\belowcaptionskip}{0pt}
\begin{center}
\vspace*{0mm}\includegraphics[width=0.6\linewidth]{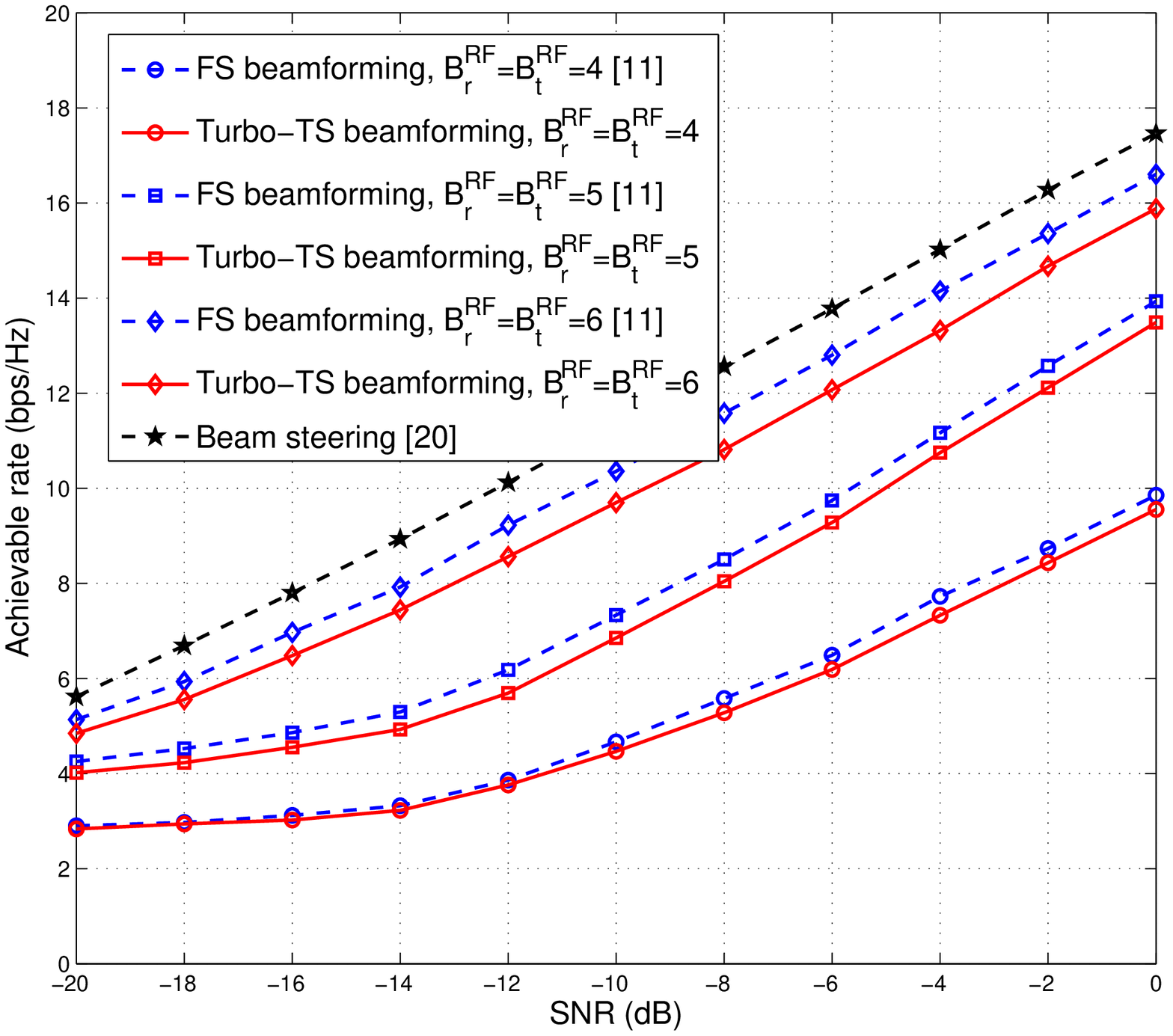}
\end{center}
\caption{Achievable rate comparison for an ${{N_r} \times {N_t} = 32 \times 128}$ mmWave massive MIMO system with ${N_r^{{\rm{RF}}} = N_t^{{\rm{RF}}} = {N_s} = 2}$.} \label{FIG4}
\vspace*{+3mm}
\end{figure}

\end{document}